\documentclass[prb,showpacs,showkeys,superscriptaddress]{revtex4}
\usepackage{amsfonts}
\usepackage{amsmath}
\usepackage{amssymb}
\usepackage{graphicx}

\begin{document}

\title{Structural behavior of uranium dioxide under pressure by LSDA+U calculations}
\author{H. Y. Geng}
\affiliation{Department of Quantum Engineering and Systems Science, The University of Tokyo, Hongo 7-3-1, Tokyo 113-8656,
Japan}
\author{Y. Chen}
\affiliation{Department of Quantum Engineering and Systems Science, The University of Tokyo, Hongo 7-3-1, Tokyo 113-8656,
Japan}
\author{Y. Kaneta}
\affiliation{Department of Quantum Engineering and Systems Science, The University of Tokyo, Hongo 7-3-1, Tokyo 113-8656,
Japan}
\author{M. Kinoshita}
\affiliation{Nuclear Technology Research Laboratory, Central Research Institute of Electric Power Industry, Tokyo 201-8511, Japan}
\affiliation{Japan Atomic Energy Agency, Ibaraki 319-1195, Japan}

\begin{abstract}
Structural behavior of UO$_{2}$ under high-pressure up to 300\,GPa has been studied by first-principles
calculations with LSDA+U approximation. The results show that a pressure induced structural
transition to the cotunnite-type (orthorhombic \emph{Pnma}) phase occurs at
38\,GPa. It agrees well with experiment observed $\sim$42\,GPa.
A new iso-structural transition following that is also predicted taking place from 80 to
130\,GPa, which has not yet been observed in experiments.
Further high compression beyond 226\,GPa will result in a metallic
and paramagnetic transition.
It corresponds to a volume of $90\,\mathrm{\AA}^{3}$ per cell, in a good
agreement with previous theoretical analysis in the reduction of
volume required to delocalize 5\emph{f} states.
\end{abstract}

\volumeyear{year}
\volumenumber{number}
\issuenumber{number}
\eid{identifier}
\maketitle


\section{INTRODUCTION}
Uranium dioxide (UO$_{2}$) is widely used as
fuel material in nuclear reactors. It was extensively
studied from 1940th in physical, chemical and thermodynamic properties by
experimental and theoretical
methods (especially during last decade) due to its important applications.
\cite{sobolev05,meis05,kudin02,pickard00,matveev97,brutzel03} The shortage of energy around the world
makes us ask for more
contribution from nuclear power, where the burn-up efficiency of nuclear fuels
is a bottle-neck. In order to tackle this difficulty, understanding the detailed
behavior of fuel materials under burning-up and irradiation is
important. Recent use of fuel
materials to high burn-up shows many microstructure formations, which is not possible to
access by empirical approach and atomic scale theoretical analysis is
highly requested.\cite{kinoshita1,kinoshita2}

Previous theoretical studies on UO$_{2}$ mainly
focused on defects effects arised from irradiation damages\cite{brutzel03,nicoll95} and thermodynamic
properties near ambient pressure\cite{sobolev05,yamada00,basak03} with semi-empirical
approaches. Electronic properties and
large scale intrinsic structural behavior under disturbance (say, compression,
tension and distortion of lattice) were rarely investigated
in despite of its importance in most properties of UO$_{2}$. This partly is due to
the lacking of a reliable method to deal with this kind of complex materials before.
The development of density functional theory (DFT)
changed the situation greatly and provides a quantum mechanics based theoretical approach to
tackle this problem.\cite{hafner00,anisimov91,anisimov93}

However, to our knowledge only a few \emph{ab initio} electronic structure studies have
been published on UO$_{2}$, most of which were based on conventional {LDA} or GGA approximation
of the exchange-correlation energy.\cite{kudin02,pickard00,boettger00,petit98,crocombette01,freyss05} It is well recognized
that strong Coulomb correlation among partly filled \emph{f} electrons of uranium
atoms makes these approximations failed. Usually a metallic ground state is
predicted for UO$_{2}$ instead of the experimental observed antiferromagnetic
semiconductor.\cite{petit96,dudarev97}
The same problem exists for transition metal oxides, and raises questions about
the applicability of DFT approach to these materials.
Fortunately, a method combining
spin-polarized local density approximation (LSDA) and on-site Coulomb repulsion
among localized \emph{d} or \emph{f} electrons,\cite{anisimov91,anisimov93} namely, {LSDA}+{U} method was
proposed and has shown its capability to treat this problem.\cite{dudarev98,dudarev00}

Usually a Hubbard Hamiltonian with two empirical parameters is employed to describe the Coulomb
interaction between 5\emph{f} electrons localized on uranium sites in UO$_{2}$.
Adding this Hamiltonian to the conventional {LSDA} (or GGS) energy functional, one
arrives at a point where all orbitals except those included in the model
Hamiltonian are treated within the framework of {LSDA} (or {GGS}) while
the localized 5\emph{f} states are treated by the unrestricted Hartree-Fock (UHF) approximation,\cite{dudarev00}
namely,
\begin{equation}
E_{LSDA+U}=E_{LSDA}[\{\varepsilon_{i}\}]+\frac{(U-J)}{2}\sum_{l,j,\sigma}\rho_{lj}^{\sigma}\rho_{jl}^{\sigma},
\label{eq:lsda+u}
\end{equation}
where $\rho_{lj}^{\sigma}$ is the density matrix of electrons occupying a partly
filled electron shell (5\emph{f} in UO$_{2}$), $\sigma$
refers to spin direction, and $\{\varepsilon_{i}\}$ is the Kohn-Sham eigenvalues.
Self-consistent solution of DFT with this energy functional gives that strong
correlation effects associated with 5\emph{f} states are going to affect all other
states as well, in particular though 2\emph{p} states of oxygen do not influenced
by Hubbard correlations directly, they are really linked to localized 5\emph{f} states
via hybridization terms. It is necessary to point out that {LSDA}+{U} method
is not a self-determined approach. The results depend on model parameters $U$
and $J$ very much, which should be chosen carefully by comparing with experimental data.
Fortunately this can be done very well with just a small set of data
and preserve the predicability of the method mostly.\cite{dudarev97,dudarev98,dudarev00}

There have been several works with LSDA+U approximation on uranium dioxide published.
All of them were near the equilibrium volume at ambiance pressure for fluorite structure
and focused mainly on electron energy loss spectra\cite{dudarev98,dudarev00,jollet97}
and magnetic structure.\cite{laskowski04} These calculations showed that
the results of {LSDA}+{U} in a well agreement with experiments. However,
no attempt was made to investigate the structural behavior of UO$_{2}$ under pressures
with LSDA+U method, which may be fundamental for
understanding the behavior of nuclear fuel under irradiations.
By far the validity of LSDA+U method beyond fluorite structure for UO$_{2}$ has not been confirmed yet.
Recent hydrostatic compression experiment\cite{idiri04} makes it
possible to check it by comparing with measured equation of state.
On the other hand, first principles calculations without Hubbard correction
on GGA(S) approximation showed that it can give almost correct energy information
for UO$_{2}$\cite{kudin02,pickard00,freyss05} regardless a wrong electronic
band structure was predicted.
Specially, by calculating the lattice parameter and bulk modulus of fluorite structure
UO$_{2}$ with various approximations, J. C. Boettger argued that density gradient
corrections, spin polarization and spin-orbit coupling effects are equally important,
and suggested when only structural properties are concerned
LSDA+U is not necessary.\cite{boettger00} However the predicted wrong ferromagnetic
ground state weakened the creditability of his argument. Other calculations ignored
spin-orbit coupling also gave reasonable lattice parameter and bulk modulus,\cite{kudin02,pickard00,freyss05}
indicates spin-orbit coupling is not so important for this case (though a large impact on
magnetic property is expected).
We will show in this paper that it should be careful when GGA(S) approximation
is used because
the coincidence
of cohesive energies of UO$_{2}$ calculated by GGS with LSDA+U approximations is valid only for
fluorite phase. An energy difference will appear if other structures are involved.

In this paper we will study the structural stability of fluorite phase (with $Fm\overline{3}m$ space group)
and cotunnite phase ($Pnma$ space group) of uranium dioxide under hydrostatic pressures
using DFT method based on LSDA and GGS approximations plus Hubbard correction.
Calculation methodology is presented briefly in next section. We will
discuss a little bit about the widely used rule of
common tangent of energy curves to determine the transition pressure of pressure-induced
structural transition, because this rule fails in a case when an energy barrier existed.
A more general rule is proposed, which can give the
energy barrier when experimental transition pressure is available.
Finally, a detailed
comparison of our results with static high-pressure experiments is given, associating with a discussion
on ultra-high pressure behavior of UO$_{2}$ crystal.

\section{METHODOLOGY}

Total energy curves of both phases ($Pnma$ and $Fm\overline{3}m$) at different volumes
are computed with VASP code.\cite{vasp,kresse96} The \emph{Pnma} structure is fully relaxed
to get all Hellman-Feynman forces smaller than $0.002$\,eV/\AA, while fluorite structure keeps
the ideal geometry due to all coordinates are completely determined by the symmetry.
For comparison, both spin-polarized generalized gradient approximation (GGS)\cite{pbe96}
and local density approximation (LSDA)\cite{lda81} with/without Hubbard $U$ term
energy functional are used. The parameters of Hubbard
term are taken as $U=4.5$\,eV and $J=0.51$\,eV, which was checked carefully
by S. L. Dudarev \emph{et al.} for fluorite UO$_{2}$.\cite{dudarev00,dudarev97,dudarev98}
Calculations employ projector-augmented wave (PAW) pseudopotentials\cite{blochl94,kresse99}
with a cutoff kinetic energy for planewaves of 400\,eV.
Integrations in reciprocal space are performed in the first
Brillouin zone with 18 irreducible k-points for fluorite structure
and at least 28 irreducible k-points for cotunnite phase generated with the
Monkhorst-Pack\cite{monkhorst76} scheme. Its convergence is well checked.
The energy tolerance for
the charge self-consistency convergence is set to $1\times 10^{-5}$\,eV for all
calculations.
Cohesive energies at different volumes
are extracted from the total energies
by subtracting spin-polarized isolated atom
contributions.
Then, they are fitted
to a Morse-type energy function
\begin{equation}
  E(V)=D\left[\left(e^{-\frac{g}{2}\left[\left(V/V_{0}\right)^{1/3}-1\right]}
-1\right)^{2}-1\right]
\label{eq:morse}
\end{equation}
to facilitate post-analysis. It is necessary to point out that for \emph{Pnma}
phase we also used a different $U$ value obtained by fitting to experimental data
of $Pnma$ phase since $U=4.5$\,eV fails
to predict the correct transition pressure. This implies that structure or lattice
distortions would have considerable impact on on-site coulomb interaction.
For the same structure, however, we find the dependence of $U$
on pressure is ignorable.

The equation of state (or compression curve) at zero-Kelvin is calculated
directly by an infinitesimal variation of cohesive energy with respect to volume given by $P=-\partial E/\partial V$.
Usually the phase transition pressure is determined by the
common tangent of their energy curves, which can be derived simply as follows.
At thermodynamic equilibrium state under finite pressure, the enthalpy must be
minimized, i.e., $\delta H=0$. In a case two phases in equilibrium, there is a variation
of enthalpy with respect to the concentration of each phase besides with respect to
volumes. The latter gives $P_{i}=-\partial E_{i}/\partial V$
(where $i$ is phase label) and the former results in
$P=-{\Delta E}/{\Delta V}={(E_{2}-E_{1})}/{(V_{1}-V_{2})}$
with $\delta H=\delta x(\Delta E+P\Delta V)$, where $\delta x$ is
the concentration variation
of, say, the first phase and $\Delta E$ ($\Delta V$) is the energy
(volume) difference between these two phases. The balance condition of pressure
requires $P=P_{i}$ (for $i=1,\,2$), namely,
\begin{equation}
  -\frac{\partial E_{i}}{\partial V}\bigg|_{V=V_{i}}=\frac{E_{2}-E_{1}}{V_{1}-V_{2}}.
\label{eq:com-tangent}
\end{equation}
It is exactly the common tangent rule for transition pressure of pressure induced
structural
transitions. Evidently, $P\Delta V$ provides the least energy $\Delta E$ required to drive a transition from
phase $1$ to phase $2$. The transition pressure equals to $P$
if no energy barrier exists, which is a
common case for usual metals and alloys. However, when an energy barrier with an amplitude of
$\Delta w$ is involved,\cite{miao03,mujica03,limpijumnong04} the work done by external pressure $P'$
should be large enough to get over the barrier in addition to the energy difference $\Delta E$.
Then the variation of enthalpy with respect to phase concentration should be
$\delta H=(\Delta E+\Delta w+P'\Delta V)\cdot\delta x\equiv 0$.
Obviously the common tangent rule becomes invalid here.
The hysteresis pressure is given by $\Delta P=P'-P$. Without knowledge about the
energy barrier, one cannot determine the transition pressure $P'$ by energy curves itself.
However, in contrast, one can deduce the energy barrier amplitude with measured
transition pressure $P'$ by
\begin{equation}
  \Delta w=-(\Delta E+P'\Delta V).
  \label{eq:barrier}
\end{equation}

\section{CALCULATIONS AND DISCUSSIONS}
\subsection{Cohesive energy}

\begin{table}
\caption{\label{tab:coheE} Cohesive energies of uranium dioxide at 0 GPa.}
\begin{ruledtabular}
\begin{tabular}{l c c c c c} 
  Phase & approach & $D$(eV/atom) & $g$ & $r_{0}$($\mathrm{\AA}$) & $B_{0}$(GPa)\\
\hline
  $Fm\overline{3}m$ & LSDA   & 9.044 & 6.122 & 5.323 & 239.99 \\
  $Fm\overline{3}m$ & GGS    & 7.956 & 6.195 & 5.432 & 203.53 \\
  $Fm\overline{3}m$ & LSDA+U & 8.194 & 6.198 & 5.444 & 208.32 \\
  $Fm\overline{3}m$ & GGS+U  & 7.212 & 6.336 & 5.552 & 180.68 \\
  $Fm\overline{3}m$ & Other calc. & 7.41\footnotemark[1]{};\;8.2\footnotemark[3]{} &  & 5.37\footnotemark[1]{};\;5.24\footnotemark[2]{};\;5.4\footnotemark[3]{} & 173\footnotemark[1]{};\;252\footnotemark[2]{};\;194\footnotemark[3]{}\\
  $Fm\overline{3}m$ & Exp.        & 7.44\footnotemark[4]{} &  & 5.46\footnotemark[4]{};\;5.473\footnotemark[5]{} &207\footnotemark[5]{};\;208.9\footnotemark[6]{}\\
  $Pnma$            & LSDA+U(\emph{U}=4.5\,eV) & 8.154 & 5.787 & 5.331 & 192.5 \\
  $Pnma$            & LSDA+U(\emph{U}=6.0\,eV)& 8.020 & 5.972 & 5.340 & 200.6 \\
\end{tabular}
\footnotetext[1]{LMTO+LSDA+U\cite{dudarev98}}
\footnotetext[2]{PW+LDA\cite{crocombette01}}
\footnotetext[3]{PW+GGA\cite{freyss05}}
\footnotetext[4]{Taken from [\onlinecite{dudarev98}]}
\footnotetext[5]{See Ref.[\onlinecite{idiri04}]}
\footnotetext[6]{See Ref.[\onlinecite{fritz76}]}\\
\end{ruledtabular}
\end{table}

Calculated cohesive energies with different approximations for fluorite ($Fm\overline{3}m$) and cotunnite
($Pnma$) structures of uranium dioxide as well as the parameters fitting to Eq.(\ref{eq:morse})
are listed in table \ref{tab:coheE}, where the energy is for per atom
and equilibrium cell volume is given by $V_{0}=r_{0}^{3}$. The cohesive
energy for a cell of U$_{4}$O$_{8}$ is given by multiplying $D$ with 12. For
comparison purpose,
other calculated and observed values\cite{crocombette01,freyss05,dudarev98,idiri04,fritz76} are also listed. It
should be noticed that \emph{Pnma} phase has a smaller effective cubic lattice constant
and bulk modulus than fluorite phase at zero-pressure, which implies it will become
stable under compression.

\begin{figure}
  \includegraphics*[0.22in,0.19in][3.9in,2.96in]{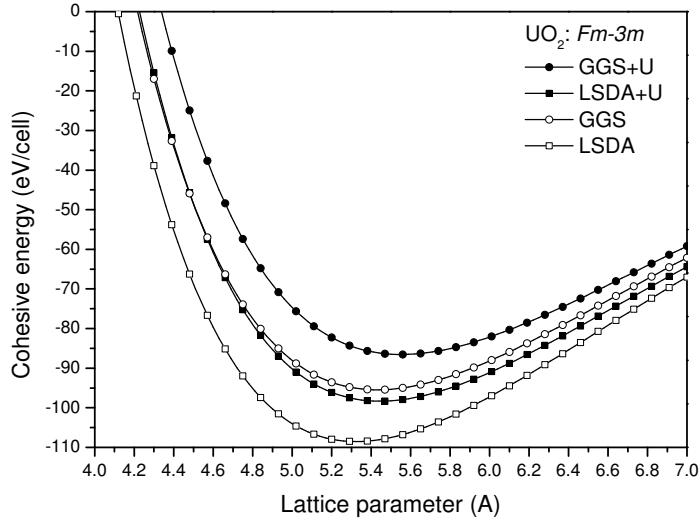}
  \caption{Comparison of cohesive energy curves for fluorite structure of UO$_{2}$
  calculated by LSDA, GGS, LSDA+U and GGS+U approximations, respectively. Notice GGS and
  LSDA+U approximations give very close energy, especially at high-compression region.
}
  \label{fig:cohE-fm3m}
\end{figure}

Variation of cohesive energy of $Fm\overline{3}m$ phase along cubic lattice constant ($V^{1/3}$)
is shown in figure \ref{fig:cohE-fm3m}. It is interesting to see
that GGS and LSDA+U give quite similar energy curves, confirming
previous calculations that GGS also can give reasonable energy information for
fluorite phase of UO$_{2}$ in spite of the corresponding electronic density
of state (DOS) is wrong.\cite{pickard00,boettger00,freyss05,note1} However,
we should emphasize here that it is just
a coincidence. Analogous to the case without Hubbard correction, LSDA+U overestimates
the binding energy slightly and GGS+U underestimates it. On the other hand \emph{U} term
uplifts the binding energy wholly, and results in this coincidence.
We can also see from table \ref{tab:coheE} that PAW method outperforms
ordinary pseudo-potentials both for GGS and LSDA approximations
in terms of equilibrium volume, cohesive energy and bulk modulus.\cite{pickard00,crocombette01,freyss05}
Furthermore, Our LSDA+U calculations with PAW potentials give results in perfect agreement with experiments\cite{dudarev98,idiri04,fritz76}
(in particular the calculated equilibrium lattice constant of 5.44\,{\AA} vs
observed 5.46\,{\AA} and bulk
modulus of 208.3\,GPa vs 208.9\,GPa). It also predicts an antiferromagnetic ground state with a
band gap of $\sim$1.45\,eV, agrees with previous calculation very well.\cite{dudarev98} To
reproduce the X-ray photoelectron spectroscopy\cite{baer80} observed band gap of $\sim$2\,eV,
Dudarev {\emph{et al}.} argued that
to take spin-orbit coupling
into account\cite{dudarev00} is necessary.
We confirmed this by a spin-orbit coupling calculation implemented in VASP which gives a band gap of 2.04\,eV.
GGS+U approximation,
however, gives a larger
equilibrium lattice constant and smaller bulk modulus, despite the cohesive energy
is more close to the observed value, as well as a band gap of 1.6\,eV.
Totally speaking, LSDA+U outperforms GGS+U approximation for this set of \emph{U} term
parameters. It is necessary to point out that the discrepancy with previous LMTO calculation\cite{dudarev98} should be owing to
their convergence precision is not so good. Their calculation gave
quite poor mechanical properties\cite{note2} that implying the force is inaccurate.
Later calculation by the same authors improved this.\cite{dudarev00}
The spin-orbit coupling is ignored in our following calculations. The resulting error
can be estimated at a lattice constant of 5.44\,{\AA} for fluorite phase, where spin-orbit
coupling decreases the cohesive energy about 0.3\,eV for per atom, very close to
the energy difference between GGS and LSDA+U at the same volume. Thus we can expect
that neglect spin-orbit coupling will gives an error of 0.012\,{\AA} in lattice constant
and 3\,GPa in bulk modulus, the same difference as GGS.

\begin{figure}
  \includegraphics*[0.22in,0.19in][3.9in,2.96in]{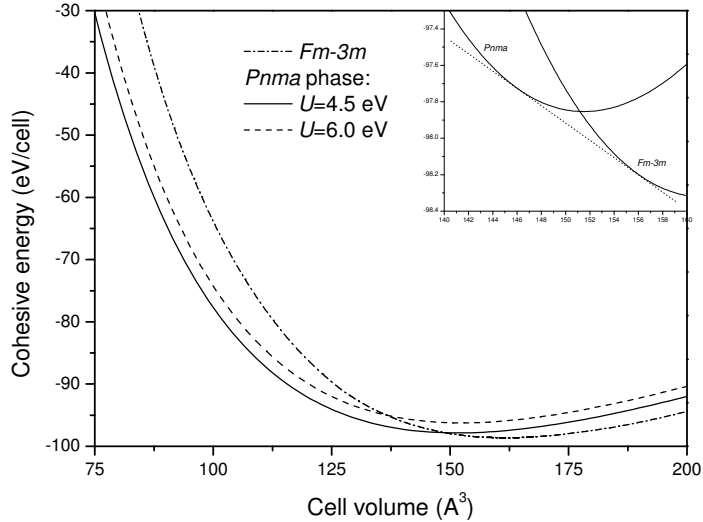}
  \caption{Comparison of cohesive energies of $Pnma$ and $Fm\overline{3}m$
  phases along cell volume. A phase transition at 7.8\,GPa is predicted by
  the slope of common tangent rule for \emph{U}=4.5\,eV case, as shown in the inset.}
  \label{fig:cohE-pnma-fm3m}
\end{figure}

\subsection{Cotunnite phase}
\label{sec:cotunnite phase}

To optimize the geometry of $Pnma$ phase at different pressure,
LSDA+U method with $U=4.5$\,eV is employed. To avoid the Pulay stress problem (which arised from the fact that the
plane wave basis set is not complete with respect to changes of the volume),
structure relaxation calculations are performed at fixed volumes rather than under constant pressures.
Then pressure is derived from the energy-volume relation.
The structure is fully relaxed to optimize all internal
coordinates and cell shape, while the symmetry of $Pnma$ space group
is kept.
Calculated cohesive energy curve is shown in figure \ref{fig:cohE-pnma-fm3m}.
For comparison the curve of fluorite phase is also given
as dash-dotted line. It shows that under high pressure $Pnma$ phase
becomes stable. A transition pressure
of 7.8\,GPa is given by the slope of common tangent as showing
in the inset. This value is quite unexpected because it is less than $1/5$
of the experiment observation as
$\sim$42\,GPa.\cite{idiri04} It is very small even if compared with another early measurement
that predicted a pressure-induced phase transition to orthorhombic $Cmcm$ phase
at $\sim$29\,GPa\cite{benedict82} (which has not yet been repeated by other authors).
Nevertheless, the calculated volume reduction of 6.4\% agrees well with the observed 7\% at the beginning
of cotunnite phase.\cite{idiri04}

\begin{figure}
  \includegraphics*[0.22in,0.19in][3.9in,2.96in]{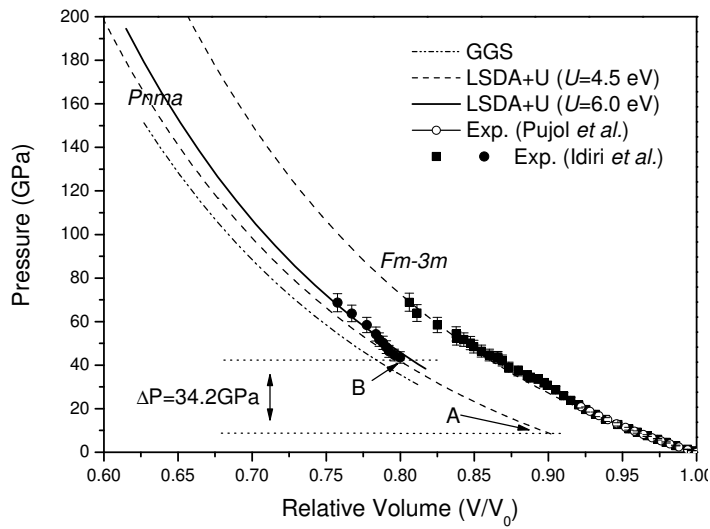}
  \caption{Calculated compression curves of uranium dioxide along relative volume compared with
  experimental measurements. An unexpected large discrepancy of
  transition pressure between measurement (point $B$) and theoretical
  prediction (point $A$) is obtained for \emph{U}45 case.
  However, a better result is recovered with \emph{U}=6.0\,eV.}
  \label{fig:EOS}
\end{figure}

Then one may ask what is the matter with it? Is the LSDA+U
approximation wrong? From table \ref{tab:coheE} and the comparison of its
results at equilibrium volume with experimental data for fluorite phase, we do not think so.
Actually, as figure \ref{fig:EOS} shows, LSDA+U gives a compression curve that
agrees very well with experiments\cite{idiri04,pujol04} for $Fm\overline{3}m$ phase,
which means that the Hubbard $U$ parameter is reasonable and insensitive to pressure.
Clearly we cannot attribute this deviation to the failure
of density functional theory or LSDA+U approximation. Figure \ref{fig:EOS} also shows the P-V curve
calculated with GGS approximation. It is worse than LSDA+U and the transition pressure
is also as low as 32\,GPa. A hysteresis pressure about 34\,GPa is estimated by
using the transition pressures observed in experiment and
calculated with $U=4.5$\,eV, which are marked by arrows B and
A in figure \ref{fig:EOS}, respectively. As discussed in previous section, this hysteresis
of transition pressure would imply
an energy barrier existing.
In fact it is very common for ionic crystal and semiconductors.\cite{miao03,mujica03,limpijumnong04} For example
a phase transition of GaN from wurtzite to rocksalt phase, where a large hysteresis
of pressure is observed.
By using Eq.(\ref{eq:barrier}), the cohesive energy curves of $Pnma$ and $Fm\overline{3}m$
phases (Eq.(\ref{eq:morse}) and parameters listed in table \ref{tab:coheE}), and
the experimental transition pressure of 42\,GPa,\cite{idiri04}
we estimate an energy barrier as $\sim$2.1\,eV per cell (U$_{4}$O$_{8}$).
This value is large enough to survive $Pnma$ phase to ambient condition.
Unfortunately, no experiment shows this event.

\begin{figure}
  \includegraphics*[0.22in,0.19in][3.9in,2.96in]{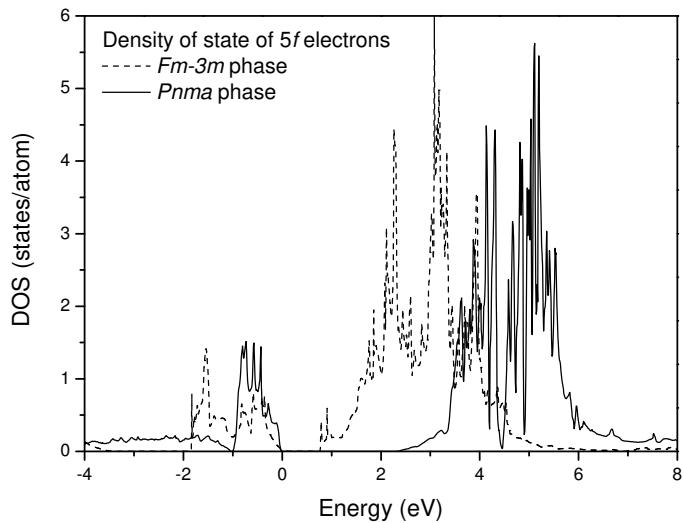}
  \caption{Electronic density of state for 5\emph{f} states of UO$_{2}$ at
  a cell volume of 131.4\,\AA$^{3}$. Transition to $Pnma$ will increase the band
  gap and shrink the energy region of localized states.}
 \label{fig:DOS-f}
\end{figure}

Therefore, the only possibility is that the discrepancy in $Pnma$ phase resulted from the dependence of
$U$ on structure (or lattice distortions). We obtained
a different $U$=6.0\,eV by fitting to the measured P-V data of
$Pnma$ phase. The resulting energy curve and Morse function parameters are
given in figure \ref{fig:cohE-pnma-fm3m} and table \ref{tab:coheE}, respectively.
We can see $U=6.0$\,eV gives a quite similar energy curve as $U=4.5$\,eV,
except the wholly uplifting of the curve. Hereafter, all calculations will be
performed for $U$=6.0\,eV and $U$=4.5\,eV separately, so we assign the former
case as \emph{U}6 and the latter as \emph{U}45 for briefness. Although the improvement on P-V curve in
\emph{U}6 is limited, as figure \ref{fig:EOS} shows, the calculated transition pressure
is corrected to $\sim$38\,GPa, almost five times of \emph{U}45 case and in a good agreement with
observed 42\,GPa. The resulted hysteresis pressure is just 4\,GPa, which ends up
an energy barrier as 0.018\,eV/atom and ignorable at room temperature. Obviously,
\emph{U}6 is more credible than \emph{U}45 since it is compatible with the fact that \emph{no}
\emph{Pnma} phase has been observed under ambient condition. The calculated
reduction of volume at transition from $Fm\overline{3}m$ to $Pnma$ phase is 6.2\%,
close to \emph{U}45 case, also agrees well with experimental data.

Figure \ref{fig:DOS-f} compares the density of state (DOS) of 5\emph{f}
states in $Fm\overline{3}m$ and $Pnma$ phases of UO$_{2}$ at a cell volume
of 131.4\,\AA$^{3}$, close to the transition pressure of \emph{U}6 case.
The most remarkable difference is the increase of band gap from 0.8\,eV
in fluorite phase to 2.4\,eV in $Pnma$ phase. As a consequence, unoccupied
states also move outwards. Below the Fermi level, different from fluorite
phase where a nearly dispersionless band containing two well-localized
5\emph{f} electrons that lies roughly from -1.8 to 0\,eV, in $Pnma$ phase
these localized states are further narrowed to start from -1.0\,eV, while the valence
5\emph{f} state is expanded from -3.7 to -1.0\,eV, too.
To completely delocalize the localized 5\emph{f}
states, a pressure above 121\,GPa is required for \emph{U}45 and
beyond 226\,GPa for \emph{U}6, we will discuss this in next subsection.

\begin{figure}
  \includegraphics*[0.22in,0.19in][3.9in,2.96in]{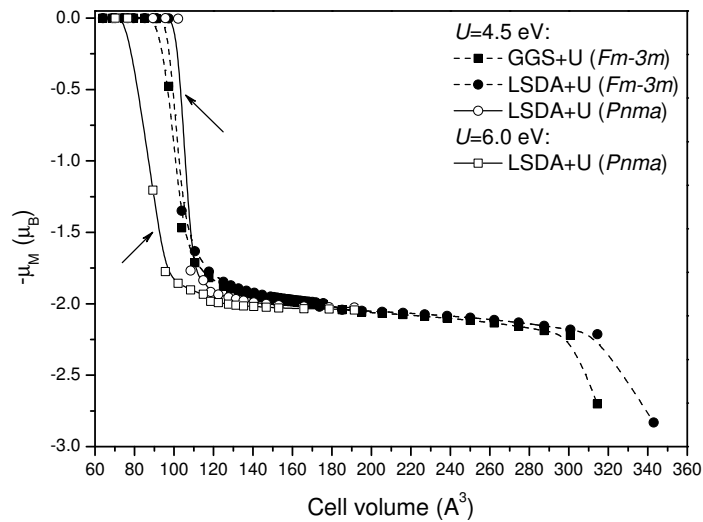}
  \caption{Variation of magnetic moment of uranium atoms with cell volume
  for $Pnma$ and $Fm\overline{3}m$ phases. The metallic transition
  is indicated by the arrows, where a paramagnetic transition also occurs simultaneously.}
  \label{fig:mag}
\end{figure}

\subsection{High pressure behavior}

The variation of local magnetic moment of uranium atoms with cell volume
is almost the same for $Pnma$ and $Fm\overline{3}m$ phases in \emph{U}45 case, implying the magnetic property is insensitive
to structural transition in UO$_{2}$. As figure \ref{fig:mag} shows,
in despite of that GGS+U and LSDA+U approximations give much different
cohesive energies for the fluorite phase, the calculated magnetic moment
of uranium atoms is very close for a large range of volume, except for
the highly expansion region ($V>300$\,\AA$^{3}$) where atoms
trend to be isolated. At equilibrium volume, our calculation gives a moment
of $\sim$1.93\,$\mu_{B}$ in a good agreement with previous
calculation\cite{dudarev98} and slightly larger than observed $1.74\,\mu_{B}$.\cite{faber76}
It can be improved by including spin-orbit coupling to $1.88\,\mu_{B}$ with
an orbit contribution of $0.46\,\mu_{B}$. This value is much smaller than
all-electron calculation where an orbit moment of $3.6\,\mu_{B}$ was predicted.\cite{dudarev00}

\begin{figure}
  \includegraphics*[0.22in,0.19in][3.9in,2.96in]{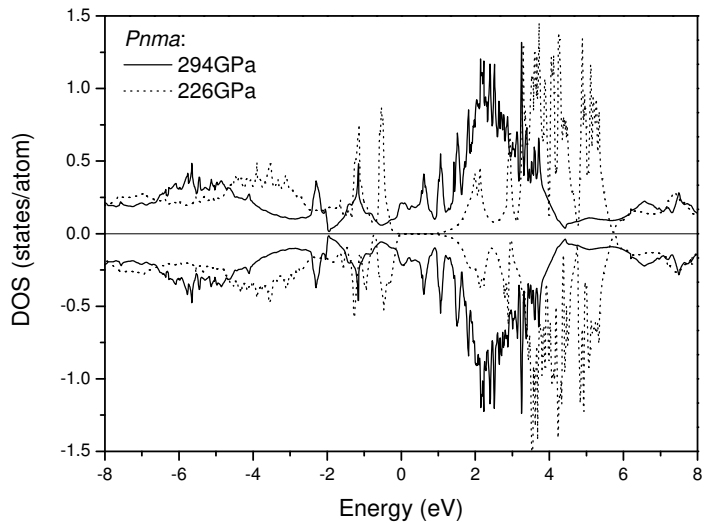}
  \caption{Total electronic density of state calculated with \emph{U}=6.0\,eV for $Pnma$ phase under
  a pressure of 226 and 294\,GPa, respectively. Here the Fermi level is at 0\,eV.}
  \label{fig:tdos}
\end{figure}

As shown in figure \ref{fig:mag}, there is a flat level for the local
magnetic moment of uranium within moderate pressure range.
A transition from antiferromagnetism to paramagnetism is observed at a volume between
102.1$\sim$108.4\AA$^{3}$ for \emph{U}45 case (equivalent to 121 and 159\,GPa in pressure).
It corresponds to a volume of 63$\sim$66.4\% of the equilibrium
volume of fluorite phase and an effective cubic lattice constant as
86$\sim$88\% of the latter phase. Increase \emph{U} to 6.0\,eV postpones the paramagnetic
transition to higher pressure as 226$\sim$294\,GPa, which has an effective cubic
lattice constant that is 82$\sim$84\% of the fluorite phase at ambient condition.
It is worthwhile to point out that at the same volume a metallic transition also occurs
due to completely delocalization of 5\emph{f} states.
Figure \ref{fig:tdos} shows the total DOS of $Pnma$ UO$_{2}$
under high pressures. We can see
the band gap disappear completely between 226$\sim$294\,GPa.
The transition volume is in a good agreement with previous intuitive
analysis that a reduction in the effective cubic lattice parameter to 82\% of the equilibrium
lattice parameter $a_{0}$ (of $Fm\overline{3}m$ phase) is required to have 5\emph{f} states in the conduction
band.\cite{kelly87} It is clear that the paramagnetic transition is driven by the delocalization of
the two pre-localized 5\emph{f} electrons which become itinerant at this volume,
and it is quite reasonable that the metallization
is always accompanied by a paramagnetic transition for materials analogous to UO$_{2}$
where both band gap and local magnetic moment are attributed to the same localized states.

\begin{figure}
  \includegraphics*[0.22in,0.19in][3.92in,2.96in]{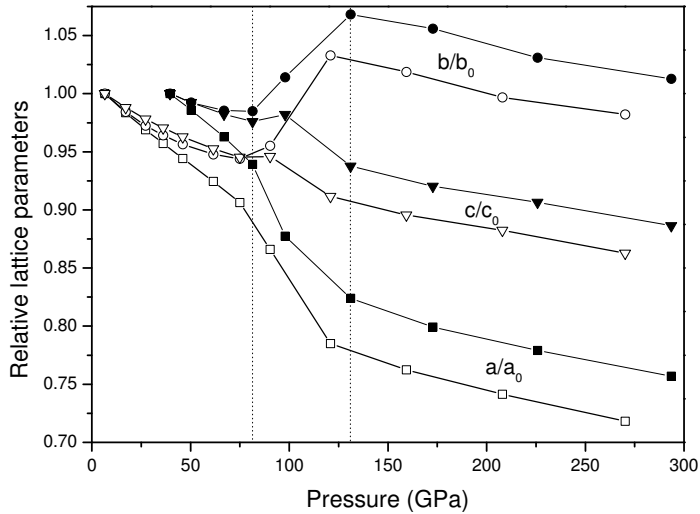}
  \caption{Pressure behavior of relative lattice parameters of $Pnma$ phase, where
  the drastic change in relative lattice constants
  (region between dotted lines) indicates an iso-structural transition.
  The curves with open symbols are calculated with \emph{U}=4.5\,eV, and those
  with filled symbols are for \emph{U}=6.0\,eV.}
  \label{fig:internal-coord}
\end{figure}

Below the metallic transition, we also find a new iso-structural transition
occurring between 80$\sim$130\,GPa for $Pnma$ phase. Figure \ref{fig:internal-coord} shows the variation of relative
lattice parameters of $Pnma$ phase starting from respective transition
pressure of \emph{U}45 and \emph{U}6 cases.
Drastic variations were observed for all lattice parameters
between 75$\sim$121\,GPa for \emph{U}45 and 80$\sim$130\,GPa for \emph{U}6,
where the smallest axis $b$ has a strong
rebound and the middle $a$ is collapsed. At higher pressure, the variations
of relative lattice parameters become smooth and approaches isotropic
compression. It is a typical structural transition.
For \emph{U}45 case one may wonder whether there is some relevancy between this
transition and the metallic one
because they adjoin closely in pressure.
However calculation with \emph{U}6 shows that they are irrelevant.
By the way, At low pressure the calculated variation
of relative lattice parameters is different from experimental observation, where
the smallest axis $b$ is most compressible whereas the $a$ axis is
most rigid. We do not know the exact reason for this discrepancy at present.
But the experiment observed trend of relative
lattice parameters cannot hold to high pressures because a stronger
repulsive force will present along the shorter axis due to higher compression of electronic states.
One can expect a rebound of the smallest axis at higher pressure.

\subsection{Intermediate structures}

As discussed in previous subsections, The value of \emph{U} depends on
structure. This raises a question about the applicability of LSDA+U
method to intermediate process of structural transition, since
the energy is affected by this term directly. It is impossible to
fit the \emph{U} value for all intermediate structures with experimental
data. Therefore, if one attempts to approximately model the transition (or lattice distortions)
with just single or several values of \emph{U},
evaluating the corresponding error becomes important. We do this job for
UO$_{2}$ by calculating the energy variation along the linear interpolated
intermediate structures
between $Fm\overline{3}m$ and $Pnma$ phases under $\sim$8\,GPa, namely,
a candidate transition path for \emph{U}45 case. In this calculation, no
structure optimizing is performed.

\begin{figure}
  \includegraphics*[0.22in,0.19in][3.9in,2.96in]{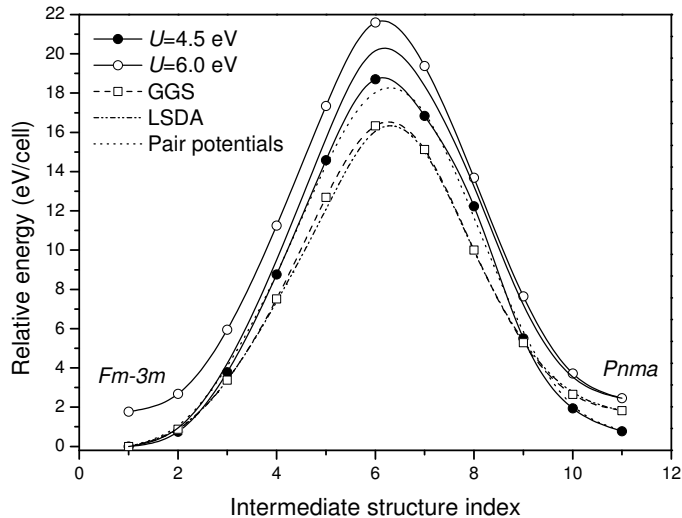}
  \caption{Relative energies of intermediate structures interpolating
  $Fm\overline{3}m$ and $Pnma$ phase linearly. The solid line without symbol is obtained
  with a linear interpolated value of \emph{U} between 4.5 and 6.0\,eV successively.
}
  \label{fig:barrier}
\end{figure}

The result is shown in figure \ref{fig:barrier}, where the respective energy of $Fm\overline{3}m$
phase is set as reference point for GGS, LSDA and a classical pair potential
model.\cite{basak03} For LSDA+U, only the energy of fluorite phase calculated
with \emph{U}=4.5\,eV is set as reference energy, to take varying \emph{U}
effect into account. Since \emph{U}6 fails to model $Fm\overline{3}m$ phase and
\emph{U}45 fails to describe $Pnma$ phase, as the first level approximation,
we interpolate the value of \emph{U} between these two phases linearly.
The result is given in figure \ref{fig:barrier} as the solid curve without symbol. As expected,
\emph{U}6 performs well for intermediate structures near $Pnma$ phase while
\emph{U}45 becomes better for those close to fluorite phase. The largest
error is 0.15 and 0.14\,eV per atom for \emph{U}6 and \emph{U}45, respectively.
What amazing is that the classical pair potential model outperforms GGS/LSDA
approximation in this test. The former has an error as 0.18\,eV per
atom and the latter two are 0.32 and 0.34\,eV per atom, respectively.
It is about two times larger than LSDA+U approximation.
This result of GGS/LSDA approximation is somewhat disappointed.
For an unit cell of U$_{4}$O$_{8}$,
it would lead to an error about 4\,eV in cohesive or formation energy.
In this sense, the point defect formation energy
calculated by M. Freyss \emph{et al}.\cite{freyss05} is inaccurate and need further
improvement with LSDA+U method in uranium defects case due to the
large structure distortions.
Finally, we would like to point out that the previous conclusion made by J. C. Boettger that ``the strong correlation effects that
are generally believed to produce the observed band gap do not have a significant
impact on the binding properties of UO$_{2}$''\cite{boettger00} should be treated carefully
depending on the studied structures and required precision.

\section{CONCLUSION}

The structural behavior of uranium dioxide under pressure up
to 300\,GPa was investigated
by DFT method with GGS/LSDA approximations plus (or not) Hubbard
\emph{U} correction for strong correlated on-site Coulomb interactions.
Comparison with experiment showed that LSDA+U gives the best
description for UO$_{2}$ in fluorite phase.
However the calculated transition pressure to $Pnma$ phase with
the same \emph{U} parameter
was quite low, indicating the value of \emph{U} depends on structure or lattice distortions
sensitively. A better value of \emph{U} for $Pnma$ phase is obtained,
which removes the factitious energy barrier predicted by \emph{U}=4.5\,eV.
The error due to varying of \emph{U} is estimated as just half of
the error given by GGS/LSDA approximation, showing LSDA+U is more reliable.
Higher pressure leads to an iso-structural transition followed by
a metallic-paramagnetic transition, which takes place
between 226$\sim$294\,GPa with an effective cubic lattice
parameter as 82$\sim$84\% of the fluorite phase's at zero pressure,
in a good agreement with previous theoretical analysis.

\begin{acknowledgments}
This study was financially supported by the Budget for
Nuclear Research of the Ministry of Education, Culture, Sports,
Science and Technology of Japan, based on the screening and counseling by the
Atomic Energy Commission.
\end{acknowledgments}


\begin{thebibliography}{99}

\bibitem{sobolev05} V. Sobolev, J. Nucl. Mater. \textbf{344}, 198 (2005).

\bibitem{meis05} C. Meis and A. Chartier, J. Nucl. Mater. \textbf{341}, 25 (2005).

\bibitem{kudin02} K. N. Kudin, G. E. Scuseria, and R. L. Martin, Phys. Rev. Lett. \textbf{89}, 266402 (2002).

\bibitem{pickard00} C. J. Pickard, B. Winkler, R. K. Chen, M. C. Payne, M. H. Lee, J. S. Lin,
J. A. White, V. Milman, and D. Vanderbilt, Phys. Rev. Lett. \textbf{85}, 5122 (2000).

\bibitem{matveev97} L. V. Matveev and M. S. Veshchunov, JETP \textbf{84}, 322 (1997).

\bibitem{brutzel03} L. V. Brutzel, J. M. Delaye, D. Ghaleb, and M. Rarivomanantsoa, Phil. Mag. \textbf{83}, 4083 (2003).

\bibitem{kinoshita1} Hj. Matzke and M. Kinoshita, J. Nucl. Mater. \textbf{247}, 108 (1997).
\bibitem{kinoshita2} M. Kinoshita, T. Sonoda and S. Kitajima \emph{et al.},
\emph{HIGH BURNUP RIM PROJECT: (III) Properties of Rim-Structured Fuel},
2004 ANS International Meeting on LWR Fuel Performance, Orlando, Florida,
September 19-22, 2004.

\bibitem{nicoll95} S. Nicoll, H. Matzke, and R. A. Catlow, J. Nucl. Mater. \textbf{226}, 51 (1995).

\bibitem{yamada00} K. Yamada, K. Kurosaki, M. Uno, and S. Yamanaka, J. Alloys. Comp. \textbf{307}, 10 (2000).

\bibitem{basak03} C. B. Basak, A. K. Sengupta, and H. S. Kamath, J. Alloys. Comp. \textbf{360}, 210 (2003).

\bibitem{hafner00} J. Hafner, Acta Mater. \textbf{48}, 71 (2000).

\bibitem{anisimov91} V. I. Anisimov, J. Zaanen, and O. K. Andersen, Phys. Rev. B \textbf{44}, 943 (1991).

\bibitem{anisimov93} V. I. Anisimov, I. V. Solovyev, M. A. Korotin, M. T. Czyzyk, and
G. A. Sawatzky, Phys. Rev. B \textbf{48}, 16929 (1993).

\bibitem{boettger00} J. C. Boettger and A. K. Ray, Int. J. Quantum Chem. \textbf{80}, 824 (2000).

\bibitem{petit98} T. Petit, C. Lemaignan, F. Jollet, B. Bigot, and A. Pasturel, Phil. Mag. B \textbf{77}, 779 (1998).

\bibitem{crocombette01} J. P. Crocombette, F. Jollet, L. Thien Nga, and T. Petit, Phys. Rev. B \textbf{64}, 104107 (2001).

\bibitem{freyss05} M. Freyss, T. Petit, and J. P. Crocombette, J. Nucl. Mater. \textbf{347}, 44 (2005).

\bibitem{petit96} T. Petit, B. Morel, C. Lemaignan, A. Pasturel, and B. Bigot, Phil. Mag. B \textbf{73}, 893 (1996).

\bibitem{dudarev97} S. L. Dudarev, D. N. Manh, and A. P. Sutton, Phil. Mag. B \textbf{75}, 613 (1997).

\bibitem{dudarev98} S. L. Dudarev, G. A. Botton, S. Y. Savrasov, Z. Szotek, W. M.
Temmerman, and A. P. Sutton, Phys. Stat. Sol. \textbf{166}, 429 (1998).

\bibitem{dudarev00} S. L. Dudarev, M. R. Castell, G. A. Botton, S. Y. Savrasov,
C. Muggelberg, G. A. D. Briggs, A. P. Sutton, and D. T. Goddard, Micron \textbf{31}, 363 (2000).

\bibitem{jollet97} F. Jollet, T. Petit, S. Gota, N. Thromat, M. G. Soyer, and A. Pasturel, J. Phys.: Condens. Matter \textbf{9}, 9393 (1997).

\bibitem{laskowski04} R. Laskowski, G. K. H. Madsen, P. Blaha, and K. Schwarz, Phys. Rev. B \textbf{69}, 140408(R) (2004).

\bibitem{idiri04} M. Idiri, T. Le Bihan, S. Heathman, and J. Rebizant, Phys. Rev. B \textbf{70}, 014113 (2004).

\bibitem{vasp} G. Kresse and J. Furthm{\"u}ller, \emph{VASP the GUIDE} (Vienna, 2005).

\bibitem{kresse96} G. Kresse and J. Furthm{\"u}ller, Phys. Rev. B \textbf{54}, 11169 (1996).

\bibitem{pbe96} J. P. Perdew, K. Burke, and M. Ernzerhof, Phys. Rev. Lett. \textbf{77}, 3865 (1996).

\bibitem{lda81} J. P. Perdew and A. Zunger, Phys. Rev. B \textbf{23}, 5048 (1981).

\bibitem{blochl94} P. E. Bl{\"o}chl, Phys. Rev. B \textbf{50}, 17953 (1994).

\bibitem{kresse99} G. Kresse and D. Joubert, Phys. Rev. B \textbf{59}, 1758 (1999).

\bibitem{monkhorst76} H. J. Monkhorst and J. D. Pack, Phys. Rev. B \textbf{13}%
, 5188 (1976).

\bibitem{miao03} M. S. Miao and W. R. L. Lambrecht, Phys. Rev. B \textbf{68}, 092103 (2003).

\bibitem{mujica03} A. Mujica, A. Rubio, A. Munoz, and R. J. Needs, Rev. Mod. Phys. \textbf{75}, 863 (2003).

\bibitem{limpijumnong04} S. Limpijumnong and S. Jungthawan, Phys. Rev. B \textbf{70}, 054104 (2004).

\bibitem{fritz76} I. J. Fritz, J. Appl. Phys. \textbf{47}, 4353 (1976).

\bibitem{note1} This is partially due to the fact that total energy is an
integral of DOS and insensitive to its detailed profile structure. A quite similar
situation holds for cluster expansion of DOS, where the convergence of DOS
is not so good, but the resulted energy and electronic entropy have a good convergce with
respect to cluster size. See, for example J. Chem. Phys. \textbf{122}, 214706 (2005).

\bibitem{baer80} Y. Baer and J. Schoenes, Solid State Commun. \textbf{33}, 885 (1980).

\bibitem{note2} From a point of view based on calculations in this paper and Ref.[\onlinecite{dudarev00}],
the criticism of J. C. Boettger\cite{boettger00} about LSDA+U method on this point is not pertinent.

\bibitem{benedict82} U. Benedict, G. D. Andreetti, J. M. Fournier, and A. Waintal, J. Phys. (France) Lett. \textbf{43}, L171 (1982).

\bibitem{pujol04} M. C. Pujol, M. Idiri, L. Havela, S. Heathman, and J. Spino,
J. Nucl. Mater. \textbf{324}, 189 (2004).

\bibitem{faber76} J. Faber, Jr. and G. H. Lander, Phys. Rev. B \textbf{14}, 1151 (1976).

\bibitem{kelly87} P. J. Kelly and M. S. S. Brooks, J. Chem. Soc. Faraday Trans. II, \textbf{83}, 1189 (1987).


\end{thebibliography}
\end{document}